\documentclass[conference]{IEEEtran}
\IEEEoverridecommandlockouts
\usepackage{cite,subfigure}
\usepackage{amsmath,amssymb,amsfonts,amsthm}
\usepackage{algorithmic}
\usepackage{textcomp}
\usepackage{booktabs} % for specialrule
\usepackage{makecell} % for diaghead
\usepackage{xcolor}
\usepackage{pgf,pgfkeys,tikz,circuitikz,float,bm,caption}
\usepackage[ruled,linesnumbered,vlined]{algorithm2e}
\usepackage{graphicx,multirow}%
\usepackage{steinmetz}
\def\BibTeX{{\rm B\kern-.05em{\sc i\kern-.025em b}\kern-.08em
		T\kern-.1667em\lower.7ex\hbox{E}\kern-.125emX}}

\graphicspath{{figures/}}

\begin{document}

\title{Doppler Spoofing in OFDM Wireless Communication Systems}

\author{ Antonios Argyriou and Dimitrios Kosmanos\\ Department of Electrical and Computer Engineering, University of Thessaly, Greece}

\maketitle

\setlength{\abovedisplayskip}{6pt}
\setlength{\belowdisplayskip}{6pt}

\begin{abstract}
In this paper we present a method that prevents an unauthorized receiver (URx) from correctly estimating the Doppler shift present in an orthogonal frequency division multiplexing (OFDM) wireless signal. To prevent this estimation we propose to insert an artificial frequency variation in the transmitted signal that mimics a transmitter (Tx) movement with a spoofed/fake speed. This spoofed Doppler shift does not affect data demodulation since it can be compensated at the legitimate receiver (LRx). We evaluate our method for its efficacy through simulations and we show that it offers a reliable way to protect one key element of the privacy of a wireless source, namely the speed of the transmitter.
\end{abstract}


\section{Introduction}
\label{section:introduction}
In recent years we have witnessed the emergence of systems that strive to localize wireless users and vehicles~\cite{Wang18b} and infer their activities~\cite{spotfi15,WiVi13,WiHear14,WiTrack14} by leveraging the wireless signal that they transmit. These are classified as passive RADAR techniques. But even though they intend to solve problems without deploying new sensing technologies they can also be used against the source of the transmission. In particular the privacy of the transmitting device (human or vehicle) is at stake since the same technology can be used by an unauthorized receiver (URx) to passively estimate speed, localize, and even infer activities without even being part of the same network. Another term for this type of problem is \textit{side-channel-information (SCI) leak}.

In this paper we strive to improve the privacy of an orthogonal frequency division multiplexing (OFDM) wireless transmitter by preventing the passive estimation of its speed. To estimate the speed of the transmitter at the URx we investigate the most reliable method which is based on estimating the Doppler shift that is added in the transmitted signal as a result of movement~\cite{Kay89,Zaihe04,Lin21}. To prevent correct estimation by the previous algorithm we construct a digitally modulated signal that has embedded an artificial Doppler shift that spoofs the transmitter speed to a desired value. Our system is designed so that the legitimate receiver (LRx) can demodulate the altered transmitted signal without compromising communication performance.

There is another observation that drives this paper: \textit{There is no fundamental reason for any receiver to know the Doppler shift introduced by movement besides removing it for data demodulation.} This is actually something that is put in practice by modern OFDM receivers that calculate the aggregate frequency offset present in a signal regardless of its origin for allowing data demodulation~\cite{moose94,beek97,Zaihe04}. More sophisticated methods can estimate just the Doppler effect in Rayleigh~\cite{Zhou08,Zheng17} and Rician fading channels~\cite{Cao17}. Even more recent methods can distinguish carrier frequency offset (CFO) between the oscillators and Doppler-induced frequency shifts in single carrier~\cite{Bellili17}, and OFDM systems~\cite{Lin21}. We tap on this last class of the most advanced systems that cannot further distinguish two different sources of Doppler. 

\begin{figure}[t]
	\centering
	\includegraphics[width=0.95\linewidth]{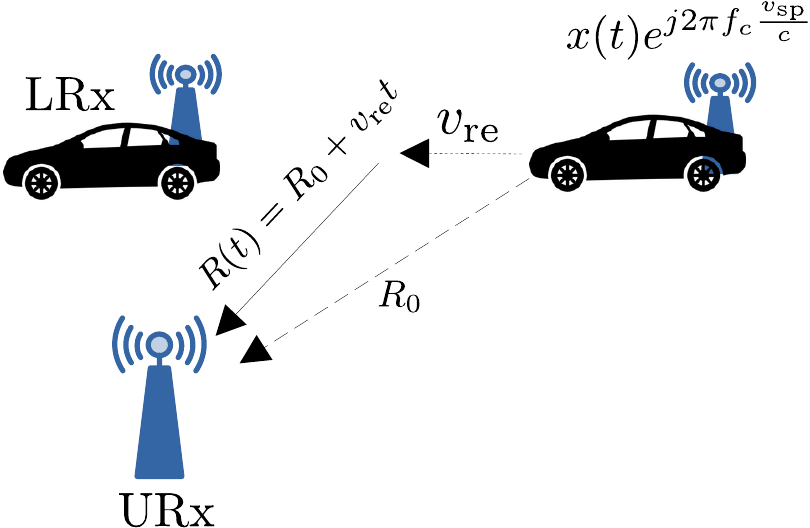}
	\caption{Scenario with a moving transmitter at speed $v_\text{re}$, and a passive unauthorized receiver (URx) that estimates Doppler. The transmitted signal contains an artificial Doppler shift of $\frac{f_c v_\text{sp}}{c}$ Hz besides the information signal $x(t)$.}
	\label{fig:car-move}
\end{figure}
It is important to note that this class of techniques is not viewed as threat to the privacy of the user. Hence, there is little focus on preventing Doppler estimation. The few existing systems focus on preventing estimation of other signal parameters. As an example friendly cryptojam~\cite{Rahbari14} prevents estimation of various PHY layer parameters but not that of Doppler. Artificial frequency shifts have been proposed in the past as a method to prevent correct demodulation by an adversary~\cite{jnl_2020_phy}, but again this method was not intended for preventing speed estimation. The most closely related work is PhyCloak~\cite{PhyCloak16} that distorts Doppler information, but it requires always the presence of a third synchronized relay node.

So how do we approach the problem of \textit{Doppler spoofing} in OFDM wireless communication? First, as the basis of our study we select an 802.11 OFDM physical layer (PHY) since most passive RADAR systems are based on it~\cite{spotfi15,WiVi13,WiHear14,WiTrack14}. Next, we study recent methods for Doppler estimation and separation from CFO in OFDM systems. We even propose an algorithm for doing the simultaneous Doppler and CFO estimation dynamically, something that is necessary for time varying channels. Second, we propose the introduction of a digital Doppler Spoofing Filter (DSF) that adds an artificial Doppler effect on the OFDM-modulated signal depending on the desired spoofed speed that the LRx should deduce. 

\section{Preliminaries} 
\label{sec:preliminaries}

\subsection{Wireless Transmission and Framing in 802.11 OFDM}
\begin{figure}[t]
	\centering
	\includegraphics[width=0.99\linewidth]{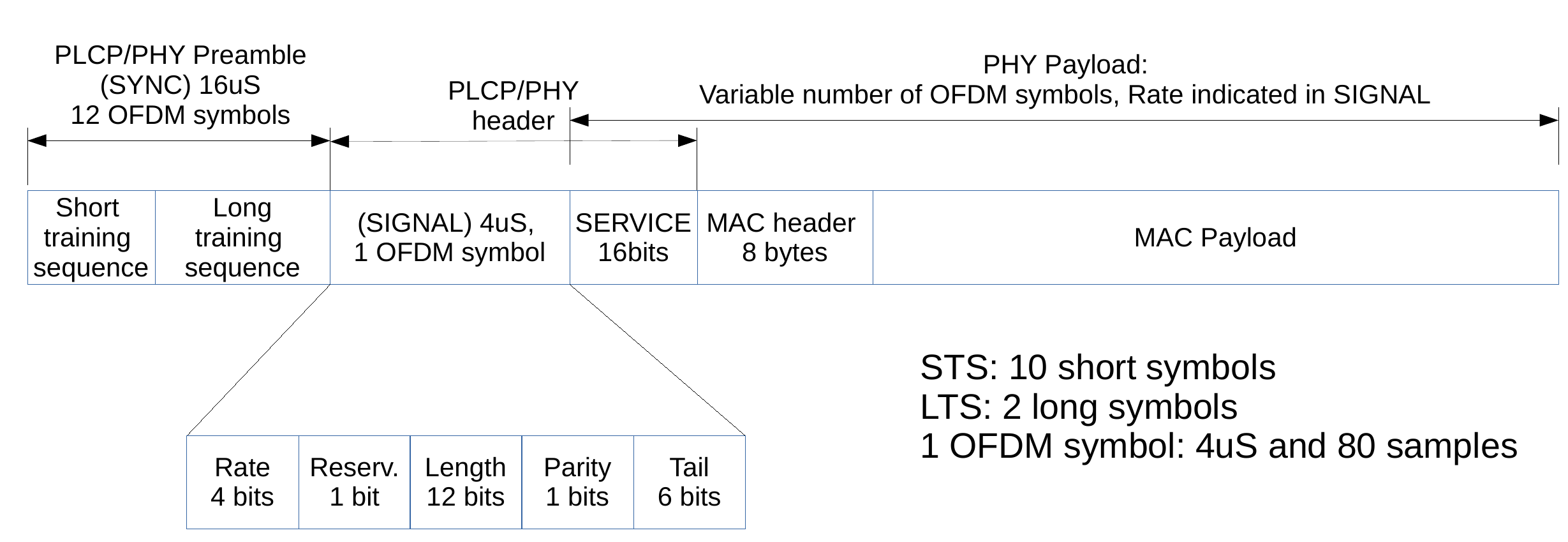}
	\caption{802.11a frame format.}
	\label{fig:wifi-frame-format}
\end{figure}

It is crucial to understand wireless frame transmission of the underlying wireless protocol, which in our case is 802.11 OFDM which was originally introduced in 802.11a. Each 802.11 frame consists of a preamble, a physical layer (PHY) header, and the MAC protocol data unit (MPDU) which are the data from the perspective of the PHY (PHY payload).
The 802.11 preamble and the associated PHY header (Fig.~\ref{fig:wifi-frame-format}) are both transmitted at the lowest supported rate (BPSK using a channel coding rate 1/2 at 6 Mbits/s). The preamble is used by the receiver for frame detection, coarse frequency estimation, and timing synchronization. It consists of 10 copies of short OFDM symbols (0.8 usec each) that constitute the short training sequence (STS), and 2 long OFDM symbols that constitute the long training sequence (LTS). The STS and LTS have both a duration of 8 usec. In terms of subcarrier utilization the STS is composed of 12 active sub-carriers modulated by the elements of a fixed sequence~\cite{yip04,80211ac}. An important note is that in this paper we use only the STS in our Doppler/CFO estimation algorithm, while more data from the LTS can be used for improving the quality of the estimates.

As shown in Fig.~\ref{fig:wifi-frame-format}, the main elements of the PHY header are the RATE and LENGTH fields, which are typically passed from the MAC. The reserved and tail bits are always zero. Once encoded, the header is sent through the rest of the transmitter  (Channel Coder, Mapper, IFFT, and Cyclic Extend) to generate the symbols. With the IEEE 802.11a/g/n/ac standard, a transmitter generates first a new random scrambling seed for every transmission of a Physical Layer  (PHY) frame (note that the header is not scrambled). The binary data are scrambled by a special construction where a 7-bit linear feedback shift register (LFSR) produces the scrambled bits. However, not all hardware vendors follow the standard. After scrambling, the channel coding module (that includes a puncturer and interleaver) encodes the data bits in a frame into redundant coded bits for robustness with a convolutional encoder of rate 1/2. Different code rates for 802.11a can be obtained by puncturing. The coded bits are mapped into a set of constellation points based on selected Quadrature Amplitude Modulation (QAM) scheme. With OFDM, these constellation points are modulated into 48 data subcarriers, while additional pseudo-random pilot symbols are modulated into pilot subcarriers for channel estimation at the WiFi receivers. The Inverse Fast Fourier Transform (IFFT) combines all these subcarriers and turns them into a time-domain signal. 

To each OFDM symbol the cyclic prefix is added to zero any inter-symbol interference (ISI). Finally, a complete WiFi frame is then upconverted and transmitted by the WiFi RF radio in a 20MHz band of either 2.4GHz(g) or 5GHz(a).

\subsection{Signal Model}
The URx may be part of the 802.11 wireless communication network of the transmitter, but is not the intended receiver of the transmission. Any station in the same 802.11 network overhears and decodes transmissions from the transmitter and during this process it is necessary to correct frequency and phase offsets (LRx in Fig.~\ref{fig:car-move}). Thus, it can indirectly estimate the Doppler and the transmitter speed.

\textbf{Complex Channel Gain:} In this paper we assume a dominant line-of-sight (LOS) path between the transmitter and the receiver that is modeled by a Rician statistical channel model. The Doppler shift is $f_D$ Hz. If the total received signal power is $P_r$ the complex channel gain coefficient is given by
\begin{equation}
\sqrt{\frac{KP_r}{K+1}}e^{j2\pi f_Dt+\phi}+\sqrt{\frac{P_r}{K+1}}g(t), \label{eqn:rician-channel-gain}
\end{equation}
where $K$ is the Rician factor that corresponds to the ratio of the power $\frac{K}{K+1}P_r$ of the LOS path versus the aggregate power $\frac{P_r}{K+1}$ from the remaining paths, and $g(t)$ is the complex Gaussian random process that corresponds to Rayleigh fading. In the high $K$ regime \eqref{eqn:rician-channel-gain} reduces to only the first LOS component. Note however, that $\phi$ is not deterministic but also a random variable (RV) uniformly distributed in $[0,2\pi]$, since it is also the result of scattering~\cite{ozdogan19}. For, easier presentation we define $h_0$=$\sqrt{P_r}e^{\phi}$ that captures the impact of the channel excluding Doppler that we will discuss next. 

\textbf{Signal Delay:} As seen in Fig.~\ref{fig:car-move} The one-way delay of the signal that arrives at the URx is time-varying and equal to $\tau(t)$=$\frac{R(t)}{c}$=$\frac{R_0}{c}-\frac{v_\text{re} t}{c}$, where $R_0$ and $R(t)$ correspond to the length of the path before the user/vehicle move and at time $t$ respectively. Now $v_\text{re}$ is the real speed that results in a path length change~\cite{Wang15} of $v_\text{re} t$ meters in $t$ seconds.

\textbf{Single-Carrier Baseband Model in Flat Fading Channel with CFO and Doppler:} The impact of CFO in a baseband signal is modeled by multiplying the signal with an exponential term of $f_\text{CFO}$ Hz. Flat fading is captured with the single complex term $h_0$ we calculated before, while the transmission delay $\tau$ simply delays the signal. If $x(t)$ represents the modulated symbol at time $t$ (e.g. QAM/PSK), and $w(t)$ is the AWGN sample the URx received signal is:
\begin{align}
	y(t)&=h_0x(t-\tau)e^{j2\pi f_\text{CFO} t}+w(t)\nonumber\\
	&=h_0x(t)e^{j2\pi(f_\text{CFO}t-f_c\tau(t))}+w(t)\nonumber\\
	&=h_0x(t)e^{j2\pi f_\text{CFO}t-f_c(\frac{R_0}{c}-\frac{v_\text{re} t}{c})}+w(t)\nonumber\\
	&=hx(t)e^{j2\pi(f_\text{CFO}t+f_c\frac{v_\text{re} t}{c})}+w(t) \label{eqn:signal-model-single-carrier-doppler-cfo}
\end{align}
In the last we set $h=h_0e^{-j2\pi f_c \frac{R_0}{c}}$. An important detail is that the Doppler shift of $f_c\frac{v_\text{re}}{c}$ Hz is related to the center frequency $f_c$, but as we will see the center frequency of the $k$-th subcarrier is $f_c+f_k$ which will make Doppler effect subcarrier-dependent. This model ignores the sampling clock offset (SCO) since we are not interested in the correct demodulation of a symbol (by sampling when the matched filter output peaks).

\textbf{OFDM Signal:} We can expand the previous model that includes the impact of Doppler when $x(t)$ is the result of wideband multi-carrier modulation and more specifically OFDM. In this case because the previous channel model considers only the LOS path it results in flat fading. With $N$ subcarriers that can contain data, pilot symbols, or a combination of both (depending on the standard), the desired time-domain OFDM symbol with duration $T_N$ seconds is produced with the use of inverse DFT (IDFT). If $f_k$ is the $k$-th subcarrier frequency, then it will be $\frac{f_k}{f_s}=\frac{k}{N}$, i.e. the ratio of the frequencies is equal to the ratio of the DFT index $k$ relative to $N$. This allows us to write the analog and the discrete (two expressions) OFDM signal that will aid us later:
\begin{align} 
	x(t)&=\frac{1}{\sqrt{N}}\sum_{k=0}^{N-1}X[k]e^{j2\pi f_kt},~~0\leq t \leq T_N.	\label{eqn:signal-ofdm-ct} \\
	x[n]&=\frac{1}{\sqrt{N}}\sum_{k=0}^{N-1}X[k]e^{j2\pi n f_k/f_s},~~0\leq n \leq N-1.\\
		&=\frac{1}{\sqrt{N}}\sum_{k=0}^{N-1}X[k]e^{j2\pi n k/N},~~0\leq n \leq N-1.
	\label{eqn:idft-ofdm} 
\end{align}
$X[k]$ is the complex QAM symbol modulated onto subcarrier $k$. The OFDM symbol consists of the $N$ samples $x[0],..,x[N-1]$.

\textbf{OFDM Signal Model:} Combining~\eqref{eqn:signal-model-single-carrier-doppler-cfo} and~\eqref{eqn:signal-ofdm-ct} gives the continuous time OFDM transmission model:
\begin{align}
	y(t)&=\frac{h}{\sqrt{N}}\sum_{k=0}^{N-1}X[k]e^{j2\pi f_k t}  e^{j2\pi(f_\text{CFO}+(f_c+f_k)\frac{v_\text{re}}{c})t}+w[n],\nonumber\\
	&~~0\leq t \leq T_N.
	\label{eqn:signal-model-ofdm-doppler} 
\end{align}

\section{Estimators}

\subsection{Joint Doppler and CFO Estimator for OFDM}
\begin{figure}[t]
	\centering
	\includegraphics[width=0.99\linewidth]{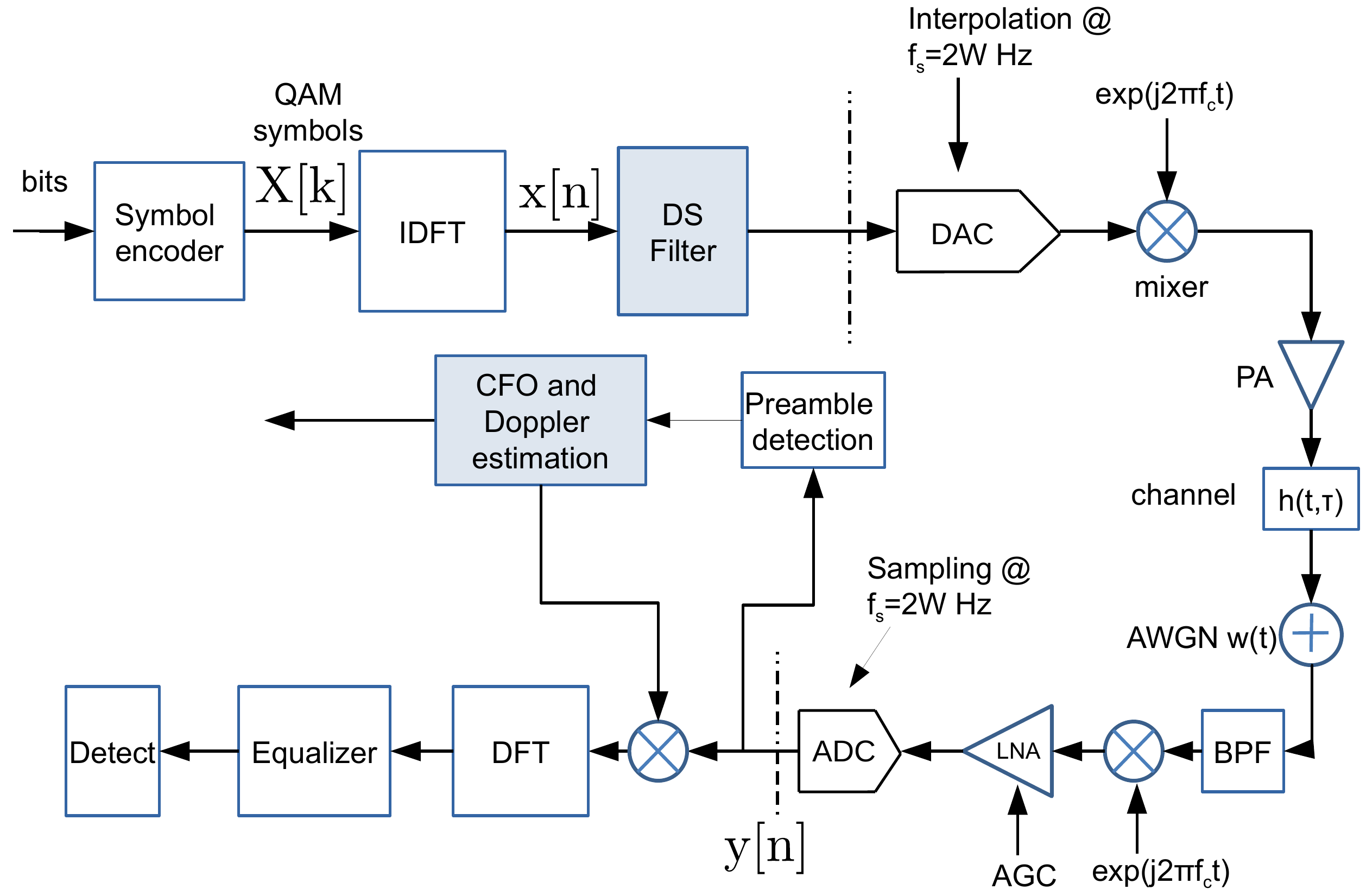}
	\caption{An OFDM transmitter with the Doppler Spoofing (DS) filter and a receiver with the dynamic joint estimator of CFO/Doppler.}
	\label{fig:doppler-spoofing-system}
\end{figure}

Not all frequency offset estimation algorithms are suitable for estimating speed since they do not differentiate the cause of the frequency shifts in the signal (CFO or Doppler). The algorithms that can differentiate leverage signal diversity across the OFDM subcarriers~\cite{Lin21,spotfi15}. Since we assume a known standard, the URx knows the location of the subcarries relative to $f_c$. So it filters the signal of each of the $N$ subcarriers that contain pilot signals in the time domain and before passing it to the DFT as illustrated in Fig.~\ref{fig:doppler-spoofing-system}. The time-domain baseband received signal for the $k$-th subcarrier is similar to~\eqref{eqn:signal-model-single-carrier-doppler-cfo}:
\begin{align}
y_k(t)&=hx_k(t)e^{j2\pi (f_\text{CFO}+(f_c+f_k)\frac{v_\text{re}}{c})t}+w(t)\nonumber \\
y_k[n]&=hx_k[n]e^{j2\pi (f_\text{CFO}+(f_c+f_k)\frac{v_\text{re}}{c})n/f_s}+w[n]
\label{eqn:signal-model-subcarrier}
\end{align}
With the known pilot $x_k[n]$ the receiver calculates the match-filtered signal for each sample $n$ of an OFDM symbol as
\begin{align}
z_k[n]=\frac{x_k^*[n]y_k[n]}{|x_k[n]|^2}.
\label{eqn:match-filter-subcarrier}
\end{align}
Then, the result for all the $N$ samples in an OFDM symbol is combined to define the statistic $U_k$ for the $k$-th subcarrier as
\begin{align}
U_k=\frac{1}{2\pi D T_s} \times  \text{Phase} \Big ( \sum^{N-1}_{n=0} z_k^*[n-D]z_k[n] \Big )
\label{eqn:matched-filter-statistic}
\end{align}
In the above $D$ is a system-selected delay between the match-filtered samples. In the previous equation we indicate the summation takes place over the duration of an OFDM symbol that consists of $N$ samples which means that the delay $D<N$, i.e. the received samples are correlated with a delayed version of themselves within the OFDM symbol. But for 802.11a/g/n/ac $D$ can be set equal to the duration of the short OFDM symbol in the STS since this symbol is repeated/delayed 10 times (see Fig.~\ref{fig:wifi-frame-format} where multiple STSs are illustrated back-to-back). In this way the symbol will be correlated with the repeated version of itself. $D$ can be selected similarly for other PHY communication schemes that use preambles with repeating symbols. 

Now by combining \eqref{eqn:signal-model-subcarrier}, \eqref{eqn:match-filter-subcarrier}, \eqref{eqn:matched-filter-statistic} and noting that when the transmitter moves with the real speed $v_\text{re}$ the Doppler shift is $\frac{v_\text{re}}{c}(f_c+f_k)$ Hz we have:
\begin{align}
	f_\text{CFO} +\frac{v_\text{re}}{c}(f_c+f_k)=U_k, \forall k \in [1,..,N]
\end{align}
Then we are able to form a linear system from the $N$ baseband signals that we have for all the used subcarriers:
\begin{eqnarray}
\left [ \begin{array}{cc}
	1 & \frac{f_c+f_1}{c}\\
	1 & \frac{f_c+f_2}{c}\\
	... & ...\\
	1 & \frac{f_c+f_N}{c}
\end{array}  \right ] \left [ \begin{array}{c}
		f_\text{CFO}\\
		v_\text{re}
	\end{array}  \right ]=\left [ \begin{array}{c}
		U_1\\
		U_2\\
		... \\
		U_N
	\end{array}  \right ]
	\label{eqn:linear-system}
\end{eqnarray}

In the above we have a number of $N$ pilot subcarriers that lead to $N$ equations and 2 unknowns ($f_\text{CFO}$ and $v_\text{re}$) making this an over-determined system that we can solve with a least squares (LS) method~\cite{Lin21}. Note that this linear system of equations is created after collecting all the $N$ samples of an OFDM symbol. So the receiver is able to estimate $f_\text{CFO}$ and the speed $v_\text{re}$ that causes Doppler. Our goal is to prevent this algorithm from correctly performing this estimation. 

\subsection{Recursive Estimator with Dynamics}
While it is not necessary for our overall system to operate (as we have explained in Section~\ref{sec:preliminaries}) $f_\text{CFO},v_\text{re}$ are assumed constant for an OFDM frame. As a result there is a set of OFDM symbols that are available for the estimation of the same parameters that is denoted as $\mathcal{M}$. Here we propose a joint recursive linear minimum mean square error (LMMSE) estimator for the problem of joint CFO and speed estimation that uses recursively these symbols. We can re-write the single measurement equation in~\eqref{eqn:linear-system} in a more condensed vector form for a single OFDM symbol $m$ that includes also the AWGN estimation error $\mathbf{w}[m]$ as
\begin{align}
	\mathbf{p}[m]=\mathbf{A}\mathbf{z}+\mathbf{w}[m],~~\forall m\in \mathcal{M},
\end{align}
where $\mathbf{z}$ is the $2\times 1$ vector to be estimated, $\mathbf{A}$ is the tall $N\times 2$ matrix in the left part of \eqref{eqn:linear-system}, $\mathbf{p}[m]$ is $N\times 1$. As we collect $N\times 1$ vector measurements $\mathbf{p}[m]$ from the m-th OFDM symbol we can combine them to refine the estimate with the recursive method. The signal model evolves over the $m$-th iteration as~\cite{kay93}
\begin{align}
	\mathbf{r}[m]=\mathbf{A}[m]\mathbf{z}+ \mathbf{w}[m].
\end{align}
We define for time instant $m$, $\mathbf{r}[m]=[\mathbf{r}[m-1]~~\mathbf{p}[m]]^T$  (a $P(m+1)\times 1$) matrix), and also $\mathbf{A}[m]=[\mathbf{A}[m-1]~~\mathbf{A}]^T$ (a $P(m+1)\times 2$ matrix). The current estimate is $\hat{\mathbf{z}}[m-1]$ and when we get the $\mathbf{p}[m]$ ($P\times 1$) vector measurement we expand the dataset $\mathbf{r}[m]$ (from $Pm \times 1$ ) to be a $P(m+1) \times 1$ vector (we accumulate more data and $\mathbf{A}[m]$ gradually becomes tall/thin). Of course $m$ can tend to infinity but in our case it will be equal to the number of available symbols in the SIFS preamble. If $\mathbf{C}_\mathbf{w}$ is the covariance matrix of the noise vector, we propose the following recursive LMMSE estimator:
\begin{align*}
	&\text{gain: }\mathbf{K}_m =\mathbf{C}_{m-1}\mathbf{A}^T(\mathbf{A}\mathbf{C}_{m-1}\mathbf{A}^T+\mathbf{C}_\mathbf{w})^{-1}~~(2\times P)\\
	&\text{estimate: }\hat{\mathbf{z}}[m]=\hat{\mathbf{z}}[m-1]+\mathbf{K}_m(\mathbf{p}[m]-\mathbf{A}\hat{\mathbf{z}}[m-1])\\
	&\text{error cov. mtx.: }\mathbf{C}_{m}=(\mathbf{I}_2-\mathbf{K}_m\mathbf{A})\mathbf{C}_{m-1}~~(2\times 2)
\end{align*}

\subsection{Artificial Frequency Shift in OFDM}
By inspecting the overdetermined system in~\eqref{eqn:linear-system} we notice that to fool this algorithm and enable estimation not only of an arbitrarily incorrect speed, but a specific spoofed speed $v_\text{sp}$ as the transmitter desires, we must insert different frequency shifts per subcarrier. For achieving the artificial speed $v_\text{sp}$ the subcarrier of each signal must experience a frequency shift of $(f_c+f_k)\frac{v_\text{sp}}{c}$ Hz, i.e. the frequency shift over different subcarriers must be different to be consistent with a moving target with a speed of $v_\text{sp}$. So to ensure a Doppler shift $(f_c+f_k)v_\text{sp}/c$ is observed, we must introduce an artificial Doppler of $(v_\text{sp}-v_\text{re})(f_c+f_k)/c$. To satisfy the previous requirement the left part of ~\eqref{eqn:linear-system} must become:
	\begin{align}
		f_\text{CFO} + \underbrace{\frac{v_\text{re}(f_c+f_k)}{c}}_\text{channel Doppler $f_D$}+\underbrace{\frac{v_\text{sp}-v_\text{re}}{c}(f_c+f_k)}_\text{artificial Doppler $f_A$}=U_k
		\label{eqn:frequency-offset-model}
	\end{align} 
Clearly we can only control the second term while we might not be able to know the channel-induced Doppler at the transmitter (i.e. know the speed $v_\text{re}$). 

\textbf{Implementation:} Regarding the implementation of our spoofing scheme note that it is not so trivial as in the case of a single frequency with a narrowband signal that we can simply multiply in the time domain with the exponential term (e.g. like~\eqref{eqn:signal-model-single-carrier-doppler-cfo}). In OFDM the frequency domain signals that serve as input to the IDFT cannot be pre-rotated since the $k$-th input to the IDFT corresponds to the multiplication with a specific exponential $e^{j2\pi n k/N}$ (see \eqref{eqn:idft-ofdm}) that itself corresponds to the "digital frequency" $k/N$ (relative to $f_s$), and $f_sk/N$ in Hz. Our approach that ensures minimal changes in existing systems is to process the OFDM symbol with the pilot subcarriers at the output of the IDFT block with a filter we name the \textit{Doppler spoofing filter} (DSF). 

\begin{figure}[t]
	\centering
		\subfigure[]{\includegraphics[width=0.98\linewidth]{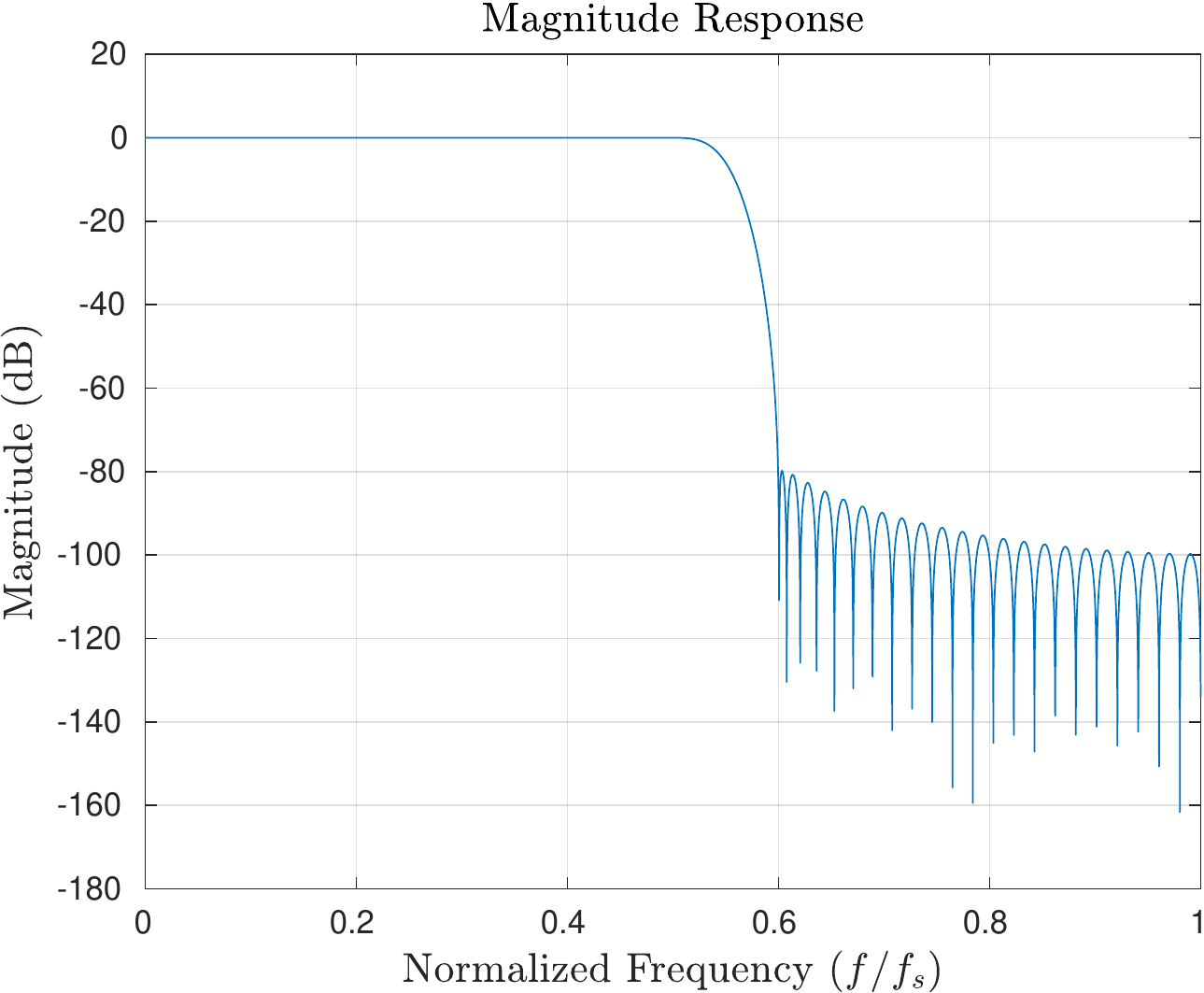}
	}\hspace{-0.2cm}
	\subfigure[]{\includegraphics[width=0.98\linewidth]{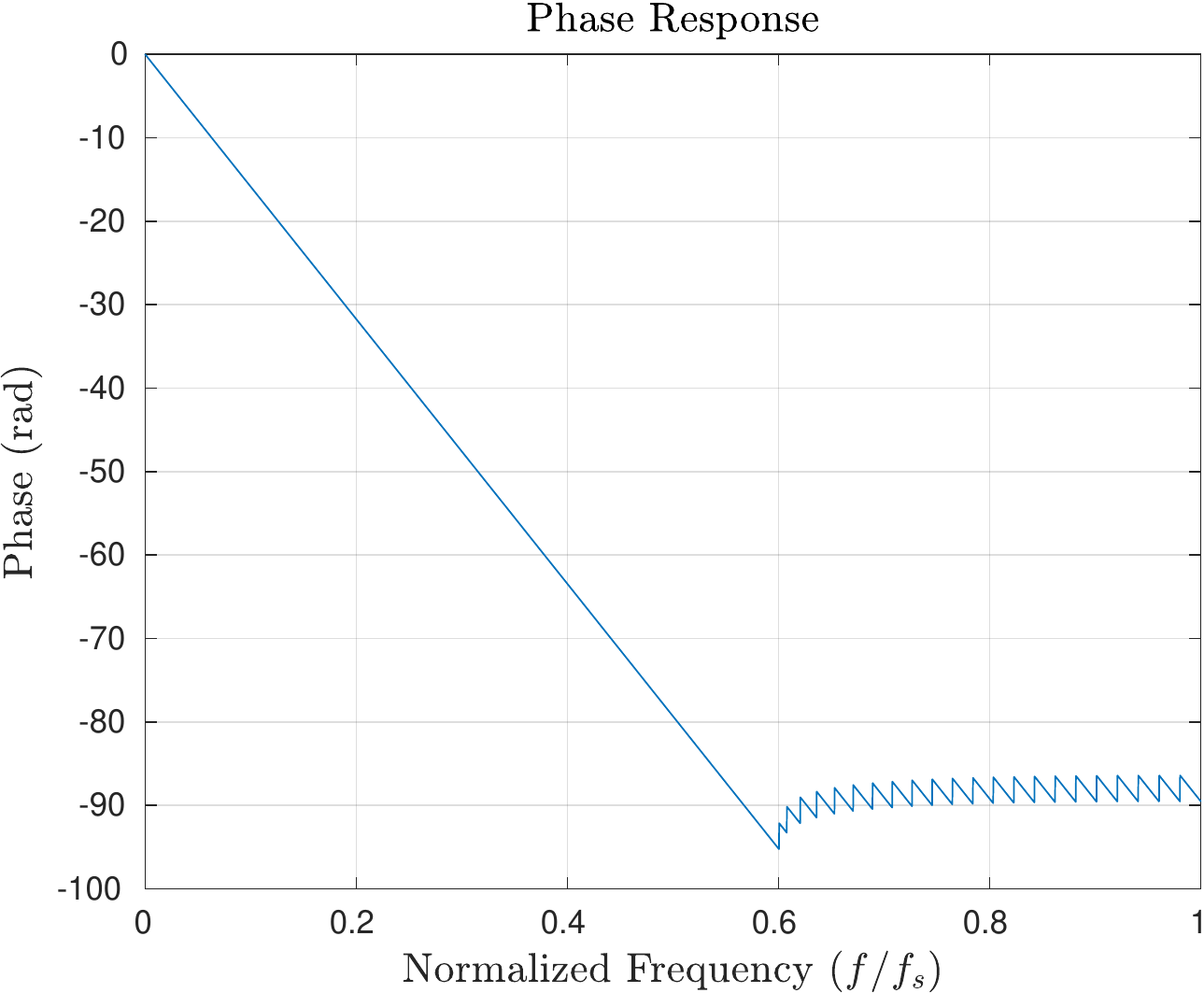}
	}
	\caption{Magnitude and phase response of the Doppler spoofing filter (DSF).}
	\label{fig:filter-tf}
\end{figure}

\textbf{Filter Design Requirements:} For a subcarrier $k$ after passing through the DSF with transfer function $H(f)$ we would like the filtered output at the receiver to be:
\begin{align}
	y_k[n]=hx_k[n]|H(f_k)|e^{j2\pi  (f_\text{CFO}+(f_c+f_k)\frac{v_\text{re}}{c})\frac{n}{f_s}+\angle{H(f_k)}}+w[n]
	\label{eqn:signal-model-subcarrier-filtered}
\end{align}
We want $|H(f_k)|$=1 and
\begin{equation}
\angle{H(f_k)}= j 2 \pi \frac{v_\text{sp}-v_\text{re}}{c}(f_c+f_k)n/f_s
\label{eqn:filter-angle-response}
\end{equation}
Thus, the filter adds a frequency-dependent delay emulating a real channel only in terms of the Doppler shift. So for baseband signals this is a lowpass filter with linear phase response. We designed a direct-form Type II FIR filter with a Kaiser window and the frequency response of the specific filter for a 802.11ac 20MHz channel can be seen in Fig.~\ref{fig:filter-tf}. Note that the phase response is completely linear over the complete passband from 0 to 0.5$f/f_s$. The sampling rate here is at 20Mhz so the stopband is at 10Mhz of the single-sideband baseband signal, while it covers the full 20Mhz of the passband signal. The location of the pilot OFDM Subcarriers in the 20 MHz channel in IEEE 802.11 can be found in the references (e.g. Fig. 2.3 of~\cite{Ghazu13}). 

\textbf{LRx Demodulation:} Inserting artificial Doppler creates inter-carrier interference (ICI) in the OFDM signal which is not desirable for the LRx. The question then is if orthogonality is destroyed on purpose can we recover it at the receiver? The answer is yes if we insert a frequency shift that is recoverable by the receiver algorithm. Similar schemes have been investigated in the literature that insert artificial frequency shift for data communication~\cite{Wang18}. In 802.11 tolerances of up to 625 KHz can be accommodated by the frequency offset estimators.

\textbf{A Simpler Scheme:} Assume that the transmitter does not know its speed. Then in \eqref{eqn:filter-angle-response} we could set $v_\text{re}$=0 and simply apply an artificial speed of $v_\text{sp}$. Then  this would add a "sub-optimal" Doppler shift in~\eqref{eqn:frequency-offset-model} since the URx would estimate the speed under the signal model:
\begin{align}
	\Delta f +\frac{v_\text{re}}{c}(f_c+f_k)+\frac{v_\text{sp}}{c}(f_c+f_k)=U_k
\end{align}
This would allow the URx to estimate the speed as $v_\text{sp}+v_\text{re}$ which is something different than the desired $v_\text{sp}$, but still it offers an option to spoof our speed as a transmitter.

\begin{figure}[t]
	\centering
	\subfigure[No Doppler shift.]{\includegraphics[width=0.95\linewidth]{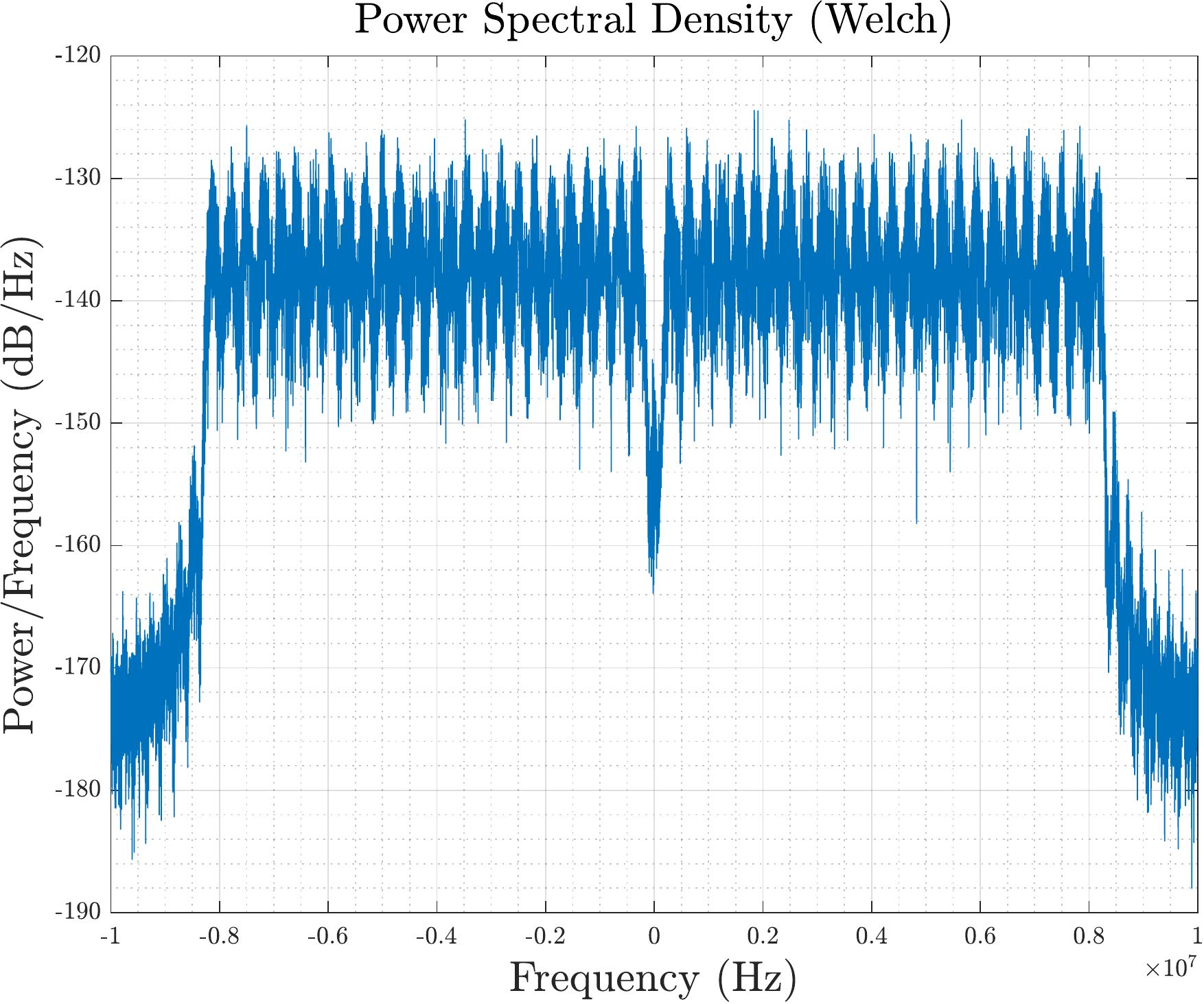}}\hspace{-0.02cm}
	\subfigure[250KHz Doppler shift on the LOS path plus 250KHz spoofed shift.]{\includegraphics[width=0.95\linewidth]{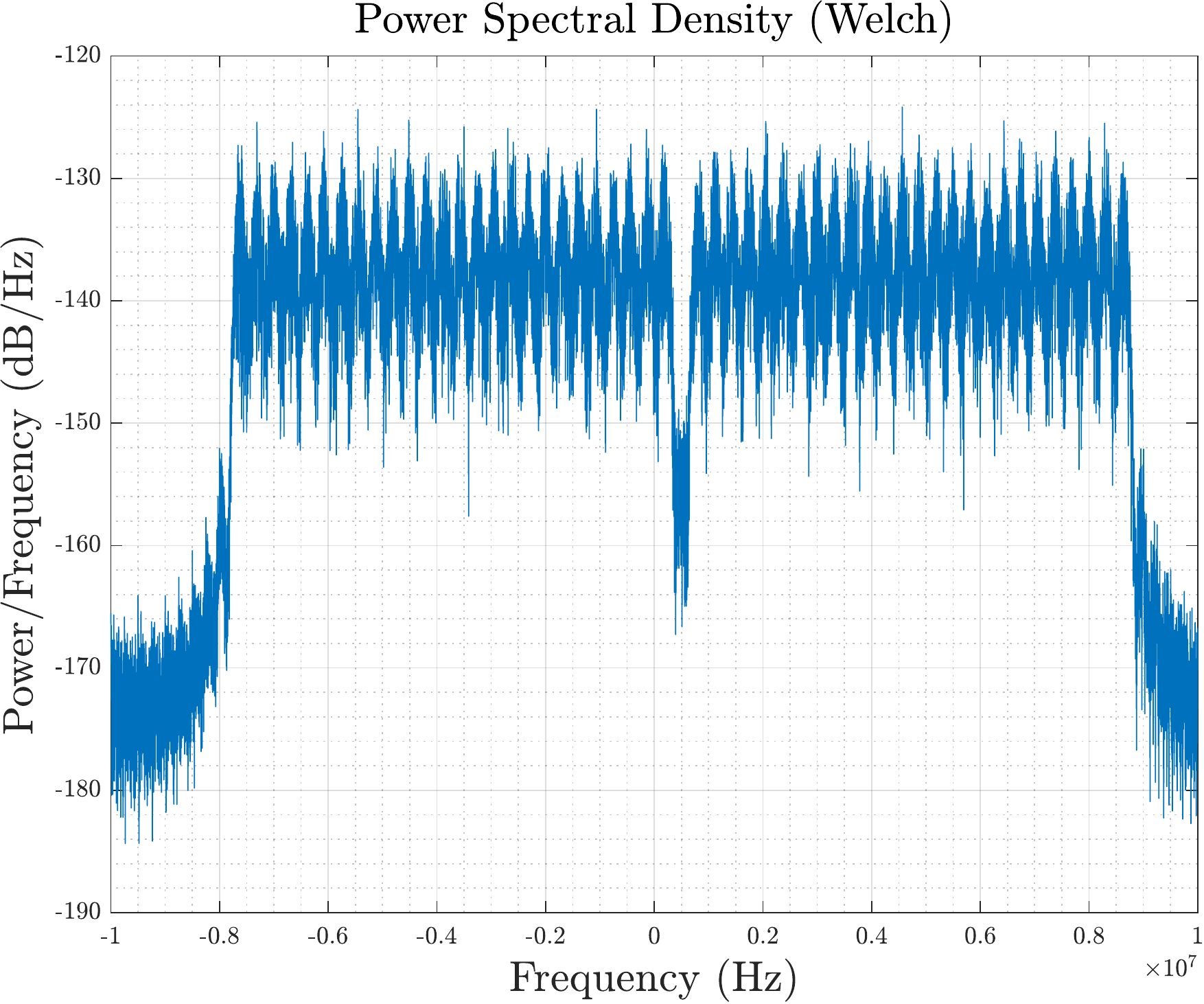}}
	\caption{20MHz baseband OFDM signal with different Doppler effects of 0 and 500KHz respectively.}
	\label{fig:periodogram2-rx-signal-ofdm}
\end{figure}

\begin{figure}[t]
	\centering
	\includegraphics[width=0.99\linewidth]{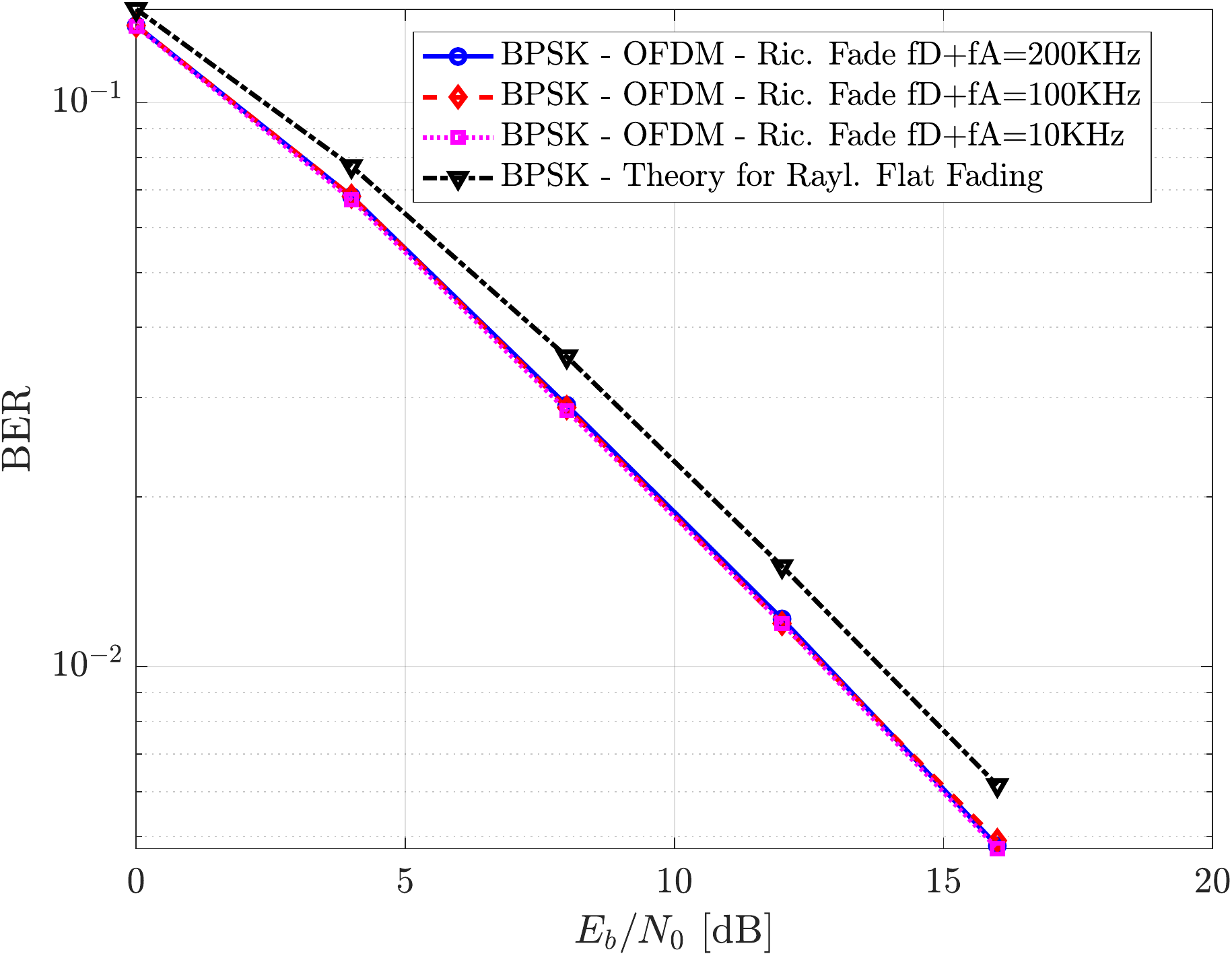}\label{fig:ber-ofdm}
	\caption{BER Results for different aggregate values of the actual Doppler frequency $f_D$ and the spoofed/artificial $f_A$.}
\end{figure}

\begin{figure}[t]
	\centering
	\subfigure[Phase estimation at the URx.]{\includegraphics[width=0.99\linewidth]{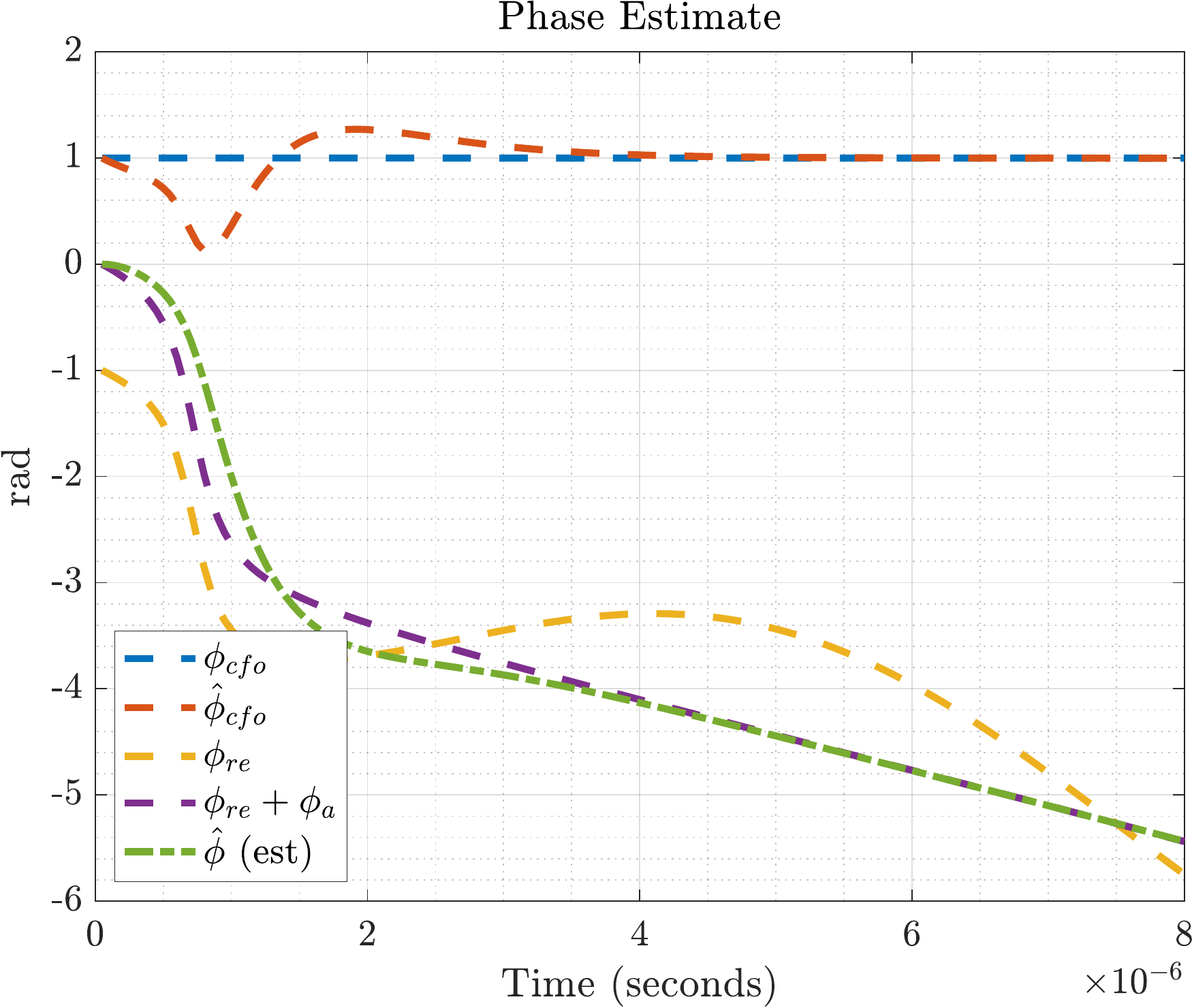}\label{fig:ber-single-carrier}}
	\hspace{-0.02cm}
	\subfigure[Average estimation error for $\phi_\text{re}+\phi_\text{a}$.]{\includegraphics[width=0.99\linewidth]{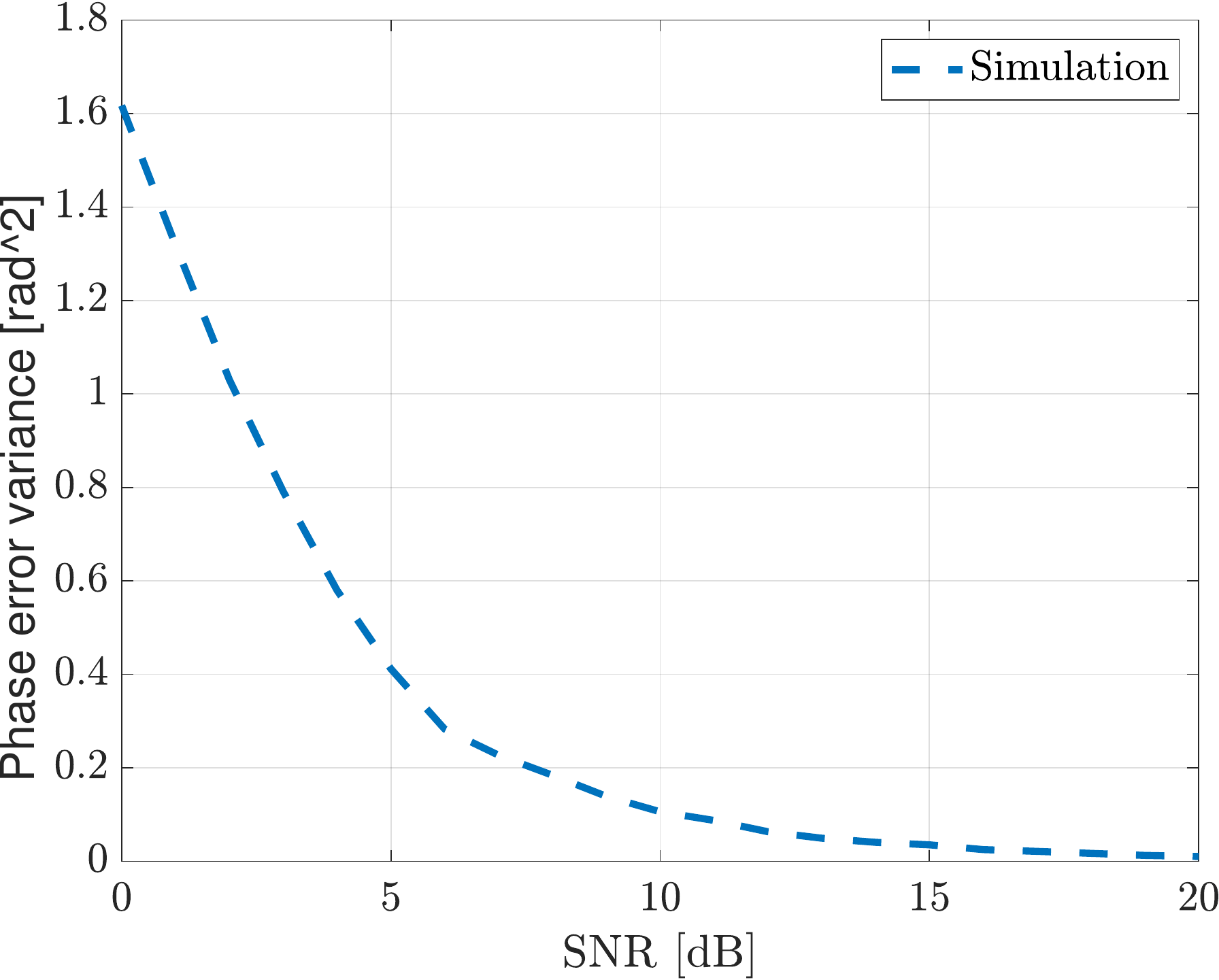}\label{fig:pll-phase-noise}}
	\caption{Estimation Results.}
\end{figure}

\section{Performance Evaluation}
The objectives of our evaluation are twofold: First to measure the performance of the LRx in terms of BER and verify that speed spoofing does not affect it. Second, measure speed estimates at the URx and verify how much it is close to the desired spoofed Doppler/speed. We considered an 802.11ac system of 52 carriers in a 20MHz channel, where 12 sub-carriers were used for pilot signals.

\textbf{Doppler Spoofing Filter Validation:} First we illustrate the performance of the DSF in terms of the periodogram of the produced signal. For large $K$, emulating thus only the LOS, and a desired spoofed Doppler of 500KHz we have the results in Fig.~\ref{fig:periodogram2-rx-signal-ofdm}. The results clearly indicate the artificial frequency shifting takes place at the desired frequency.

\textbf{BER Performance:} Fig.~\ref{fig:ber-single-carrier} presents the BER performance of the LRx over the Rician fading channel when it uses the Moose algorithm~\cite{moose94} which adopts established well-known principles for synchronization in OFDM receivers. Note that the LRx has no reason to deploy the joint CFO/Doppler estimation algorithm developed in this paper since it only needs to find the overall phase/frequency shift and remove it for demodulation.

\textbf{Doppler Estimation:} We present the result of real-time Doppler estimation vs. time in terms of the estimated phase in Fig.~\ref{fig:ber-single-carrier}. First, we see that it can estimate successfully the CFO which due to its small value does not create large phase variation over the time of 8us that a SIFS lasts. We notice that the remaining results are as expected, that is the receiver can estimate from the 160 available samples of the SIFS the time-dependent phase of the signal $\phi=\phi_\text{re}+\phi_\text{a}$ that includes the real Doppler and artificial Doppler. At the same time we present the actual time-dependent phase of the signal $\phi_\text{re}$ due to Doppler in the Rician channel which also reveals the speed. The inability of the system to separate the source of real channel-induced Doppler shift is evident. Besides the real-time results, we also present in Fig.~\ref{fig:pll-phase-noise} the average error for the estimation of $\phi_\text{re}+\phi_\text{a}$ which presents an improving trend for high signal-to-noise ratio (SNR) in the received signal. Still, the previous quantity is estimated as a single parameter by the algorithm at the receiver regardless of the SNR.

\section{Conclusions}
In this paper we presented a new approach for improving the privacy of OFDM wireless communication by preventing Doppler estimation by an unauthorized receiver. The basic idea suggests the insertion of an artificial (spoofed) frequency variation at the transmitted signal that disallows Doppler estimators to produce correct estimate of the channel-induced Doppler, and eventually of the speed. Our results indicate that modern joint CFO/Doppler estimation techniques (including our variation for a dynamic system) cannot separate the source of the Doppler effects in the signal, improving thus the privacy of the transmitter by preventing estimation of the Doppler effect induced by its speed.

\bibliographystyle{IEEEtran}
\bibliography{../../../../../tony-bib}

\end{document}